# Correlation structure of STAR events[1]

Mikhail Kopytine for the STAR Collaboration
Department of Physics, Kent State University, Kent, Ohio, USA

**Abstract.** STAR observes a complex picture of RHIC collisions where correlation effects of different origins -- initial state geometry, semi-hard scattering, hadronization -- coexist. I present STAR measurements of particle number and velocity field correlations obtained using a variety of techniques, including direct construction, fluctuation inversion and discrete wavelet transform. Connections between the techniques are discussed in detail. The resulting physics picture is presented in the context of identifying the relevant degrees of freedom and the likely equilibration mechanism.

## 1. Introduction

This talk is a partial overview of recent STAR results in the field known as "correlations and fluctuations". Out of multiple reasons to analyze correlations and fluctuations at RHIC known before the data became available, there is one that seems to dominate now: by measuring correlations and their evolution we learn about equilibration of the system. Is it taking place? What is the mechanism? And *what* is equilibrating? Novel and advanced data analysis techniques have been created lately to address these issues. I will focus on the developments since Quark Matter 2004.

## 2. Definitions

In central AuAu events reconstructed by STAR, the multiplicity of charged tracks may be as high as $10^3$. When a particle is characterized by a quantity $x$, a pair of particles is characterized by a respective pair $(x_1, x_2)$. In general, the two-particle correlation analysis can be carried out in a space which doubles the number of dimensions used for the single-particle analysis. The object of analysis is the differential density $\rho$ which characterizes the distribution of the quantity of interest (be it particle number or another kinematic variable) in this space. We distinguish sum and difference variables $x_\Sigma$ and $x_\Delta$:

$$\begin{pmatrix} x_1 \\ x_2 \end{pmatrix} \rightarrow \begin{pmatrix} x_\Sigma \equiv x_1 + x_2 \\ x_\Delta \equiv x_1 - x_2 \end{pmatrix} \quad (2.1)$$

---

[1] This document includes copyrighted material to be published in International Journal of Physics G, © 2005, IOP Publishing Ltd.

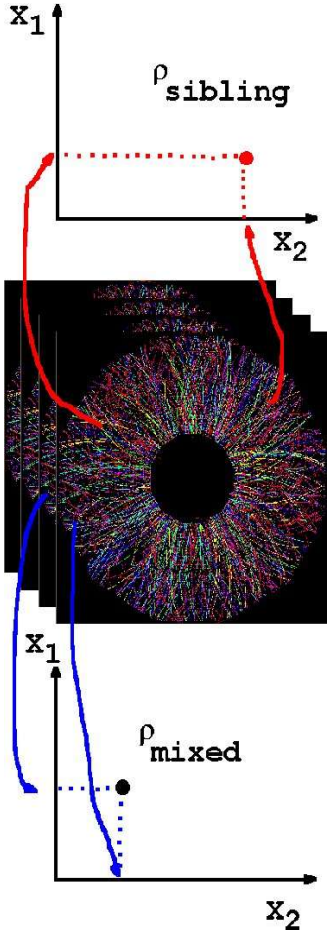

**Figure 2.1** Construction of sibling and mixed pair densities $\rho$ in the space of kinematic variable $x$ by histogramming.

In many cases of interest, the $x_\Sigma$ dependence is absent or negligible. Then, $\rho(x_1, x_2)$ or equivalently $\rho(x_\Sigma, x_\Delta)$ can be reduced to $\rho(x_\Delta)$ without loss of information. The latter quantity is often (especially in the fields of study where $x$ is time) called *autocorrelation* function. Joint autocorrelation extends the same concept from one-dimensional $x$ to the case of a two-dimensional space such as $(\eta, \phi)$.

## 3. Charge-dependent number correlations– modified hadronization in the medium

The charge-dependent (CD) number correlation is defined as a difference between the like-sign and the unlike-sign correlations. (In this and subsequent sections we discuss results obtained without particle identification). The correlations are defined by comparing sibling (same-event) pairs with mixed-event pairs, obtained from mixing similar events. To characterize and normalize correlations, we use

$$\frac{\Delta \rho}{\sqrt{\rho_{ref}}} = \frac{\rho_{a,b} - \rho_a \rho_b}{\sqrt{\rho_a \rho_b}} \quad (3.1)$$

where $a$ and $b$ index particles (or in the discretized case, bins in the kinematic space occupied by particles), $\rho_{a,b}$ is pair density distribution, while $\rho_a$ and $\rho_b$ are the corresponding single-particle distributions, whose product forms the reference density $\rho_{ref}$. Experimentally one deals with fluctuating histogram bin contents $n_a$ and $n_b$ and Eq.3.1 becomes

$$\frac{\Delta \rho}{\sqrt{\rho_{ref}}} \rightarrow \frac{\overline{n_a n_b} - \overline{n_a}\,\overline{n_b}}{\sqrt{\overline{n_a}\,\overline{n_b}}} = \frac{Cov[n_a, n_b]}{\sqrt{\overline{n_a}\,\overline{n_b}}} \quad (3.2)$$

If the variance in the bin population $n$ is dominated by the Poissonian process,

$Var[n] = n$, and if in presence of correlations $Cov[n_a; n_b] \propto Var[n]$ (number of correlation sources is proportional to multiplicity) this normalization eliminates the trivial multiplicity dependence of the correlation amplitude. Eq.3.1 represents a per-particle measure of correlations. This means that under such assumptions, one can directly compare results obtained from events of different multiplicity as long as Eq.3.1 is used.[2]

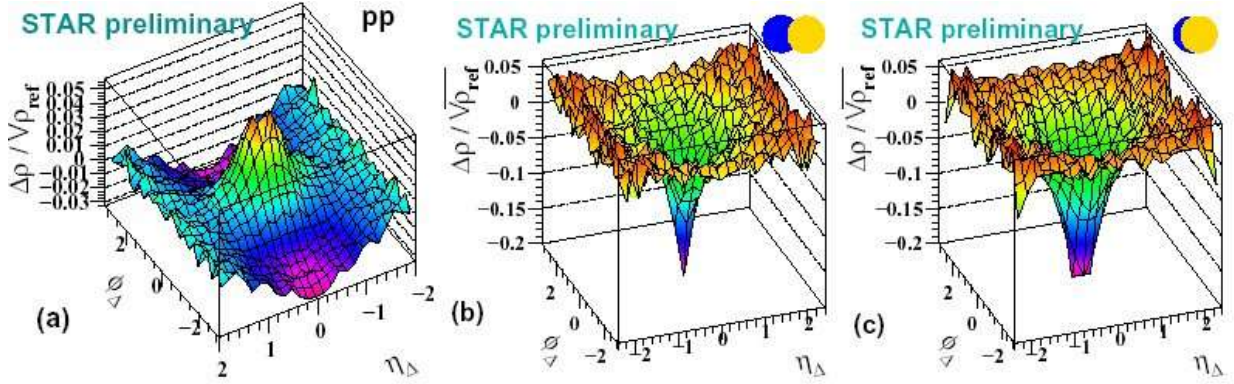

**Figure 3.1** Comparison of CD correlations in pseudorapidity and azimuthal angle: **a)** pp collisions at $\sqrt{s}$ =200 GeV, **b)** peripheral AuAu collisions, **c)** central AuAu collisions. **b)** and **c)** are at $\sqrt{s_{NN}}$ =130 GeV.

In paper[1], CD correlations in η and φ have been analyzed simultaneously by using *joint autocorrelation*. Figure 3.1 shows a comparison of CD correlations in pp collisions (panel a) with peripheral and central AuAu collisions. Kinematic cuts on the track separation and a $p_t$ cut are applied to eliminate the contribution of HBT and Coulomb effects in AuAu collisions (panels **b** and **c**)[1]. A much broader HBT correlation for pp collisions contributes the Gaussian peak at the origin in **a**.

Transverse momentum conservation also contributes to the structure at the origin (the ridge beneath the Gaussian peak). The main feature, the negative Gaussian peak on $\eta_\Delta$, is attributed to local charge conservation during hadronization. If the process happens on a longitudinally expanding string (as e.g. in Lund model), the strict alternation of positive and negative hadrons in rapidity holds (modulo interspersed neutrals). This implies that $\eta_\Delta$ for an unlike-sign pair tends to be shorter than for the like-sign, regardless of $\phi_\Delta$, creating a groove in the CD plot, clearly seen in Fig.3.1-**a.** The $\phi_\Delta$-independent groove appears to diminish in magnitude from **a** and **b** to **c**. In addition, the negative peak at 0 relative angle changes its shape, becoming narrower from peripheral to central AuAu collisions and acquiring more $\eta - \phi$ symmetry. We hypothesize that this reflects a change in the hadronization geometry from one dominated by independent string break-up in peripheral collisions to a bulk hadronization process in which individual strings are no longer relevant.

---

[2] In this terminology, the correlation function commonly used in HBT represents a per-particle-pair-normalized quantity. The choice of normalization depends on the correlation mechanism under study.

## 4. Charge-independent number correlations – longitudinal minijet broadening

The charge-independent (CI) number correlation sums the like-sign and the unlike-sign correlations. In AuAu collisions, these correlations have been analyzed by STAR in the space of pseudorapidity and azimuthal angle [2, 3] and in the transverse momentum space [4]. Our discrete-wavelet technique [2] extracts information about correlation structure by measuring the power spectra of local fluctuations in the density of charged hadrons with respect to a mixed-event reference (the so-called "dynamic texture" of the event). As discussed in Appendix B, this observable is a measure of the gradient of the two-particle correlation function, which bypasses construction of a correlation function. The experiment [2] reveals a reduction in the dynamic texture (and thus in the correlation function gradient) along the $\eta$ -direction at $p_t > 0.6$ GeV/c in the central AuAu collisions, compared to an expectation based on peripheral events. HIJING simulations [2] point to minijets as the likely source of correlations of this scale and magnitude. The possible mechanism of the reduction is a coupling between the minijet fragments and the longitudinally expanding bulk medium. In paper [3], the same correlation structure is investigated in the joint autocorrelation technique. It is found that the correlations are broadened in $\eta$ with centrality. A reduction in the correlation gradient ("dynamic texture") and the elongation of the correlation function along $\eta$ are consistent descriptions of the effect.

In paper[4], particle number correlaitons in $p_t$ have been studied in AuAu collisions at $\sqrt{s_{NN}}$ =130 GeV. With increasing collision centrality, the correlation is found to distribute itself more evenly on $p_t$, depleting its higher $p_t$ part (associated, again, with semi-hard scattering).

## 5. Transverse-momentum correlations – minijets and elliptic flow

We require a direct measurement of correlations in the velocity field, independent of particle number correlations. To form an independent measurement, these have to be separated from particle number correlations. For a random field (such as particle number or velocity distributed in $\eta$ and $\phi$, see Fig.5.1) variance of the integrated content of a bin can be related via an integral equation to the correlation function of the field [5, 6]. The correlation function can be reconstructed from the measured variance by solving the integral equation derived in [5, 6]. This process offers a computational advantage since the bin integration involves O($N$) computations, whereas two-particle correlations require O($N^2$) ($N$ is the event multiplicity used). By varying the bin size one varies the range of the difference variable in the correlation function.

In order to extract the $p_t$ correlation alone, we need to disentangle the effect of $n$ from that of $p_{t,i}$ in the variance of $p_t \equiv \sum_{i \in (\eta,\phi)bin} p_{t,i}$, and from statistical fluctuations. We denote the desired variance as $\Delta\sigma^2_{p_t:n}$. As an estimator of

$\Delta\sigma^2_{p_t:n}$, we use $Var[p_t - n\hat{p}_t]/\bar{n} - \sigma^2_{\hat{p}_t}$, where the latter term is the inclusive per-hadron $p_t$ variance. In the case when p_t in the bin, event-by-event, derives from the same parent distribution (not the case in Fig.5.1), $\Delta\sigma^2_{p_t:n}$ converges to 0, despite

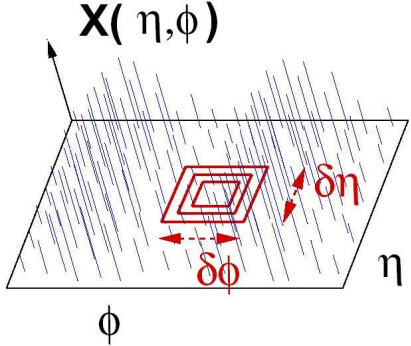

**Figure 5.1** A fantasy event. Positions of spikes indicate $\eta, \phi$ kinematics of individual particles, their heights – magnitude of quantity X assigned to a particle, which could be $p_t$ or a derived quantity. This event has azimuthal anisotropy, however its $v_2$ may be 0, illustrating the difference between number and $p_t$ flow. The analysis proceeds by selecting a bin of size $\delta\eta, \delta\phi$, integrating the random field $X$ within the bin, finding the variance of that integral in an event sample, then feeding this information into an integral equation to yield the correlation of $X$, as discussed in Appendix B.

fluctuations (statistical or not) in $n$.

Top panels in Fig.5.2 represent scale dependence of the variance $\Delta\sigma^2_{pt:n}$ for three centrality classes. The bottom panels, representing per-particle normalized correlations plotted on difference variables, are obtained from the respective top panels by solving the integral equation derived in [5, 6] (see Appendix C) which relates variance and correlation. While the quadrupole wave in Fig.5.2 (bottom row) is the reflection of the elliptic flow in the $p_t:n$ correlation function, the peak at 0 angle difference likely reflects minijets[7]. The amplitude of the minijet peak has a non-monotonic behavior with increasing centrality, rising in peripheral and falling in central collisions. The peak (Fig.5.3) is seen to be elongated longitudinally in central collisions, showing qualitatively the same behavior as seen with number correlations by using wavelet analysis and joint autocorrelation (Section 4).

## 6. Conclusions

I conclude with a broad-brush sketch of the physics picture inferred from these "soft" $p_t$ (<2 GeV) measurements. The *important* degrees of freedom are the *collective* ones manifesting themselves through correlations of "ordinary" hadrons. In the charge-independent sector, these degrees of freedom are associated with the motion of numerous fragmenting objects, which we hypothetically identify with partonic systems having suffered semi-hard scattering. Due to kinematic constraints in the course of the fragmentation, they transfer the correlation from higher to lower $p_t$ in a process which is found to be centrality-dependent. In particular, the phenomenon of $\eta$-broadening of minijets at "soft" $p_t$ seen with both number and $p_t$ correlations (Sections 4 and 5) can clarify the nature of the medium created in central

AuAu collisions, and of its equilibration mechanism, provided that the coupling mechanism is under theoretical control. Increased symmetry of the charge-dependent correlations on ($\eta$, $\phi$) in the central collisions (Section 3) may point to a change in the hadronization geometry in the medium.

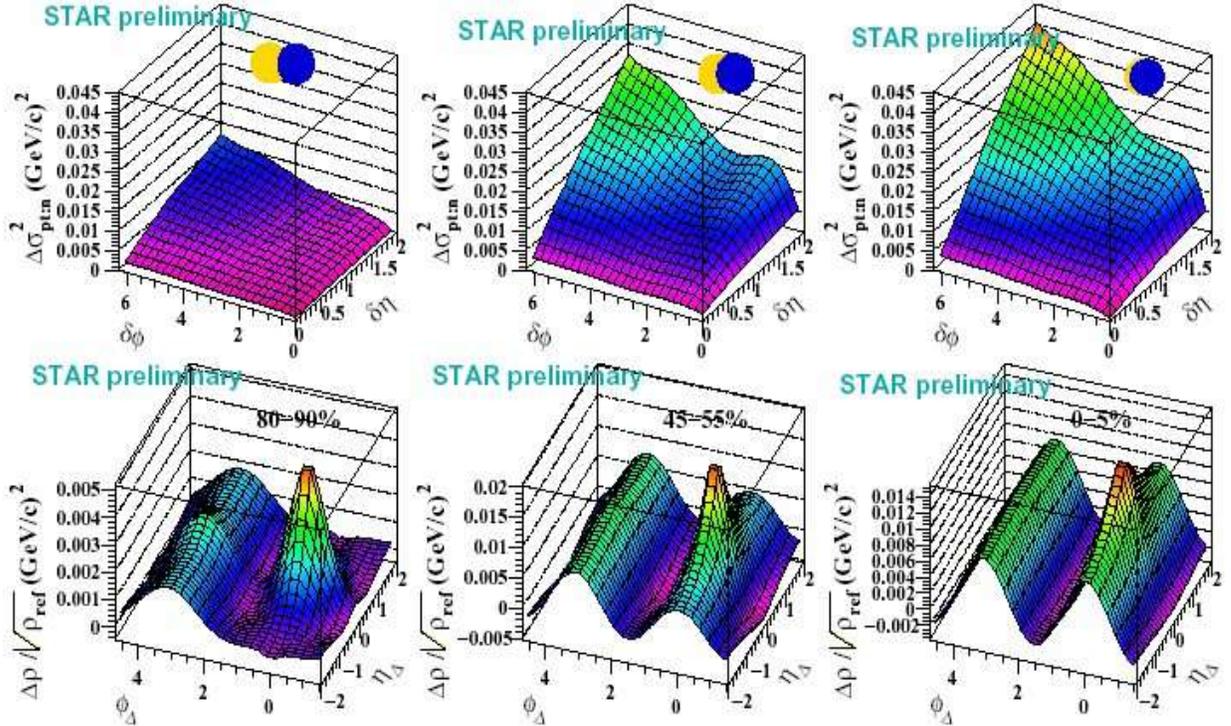

**Figure 5.2** Top: scale dependence of the reference-subtracted variance $\Delta\sigma^2_{p_t:n}$. Bottom: reference-subtracted normalized correlation functions reconstructed from $\Delta\sigma^2_{p_t:n}$ by inversion. The percentage intervals indicate centrality for each column-wise pair of plots.

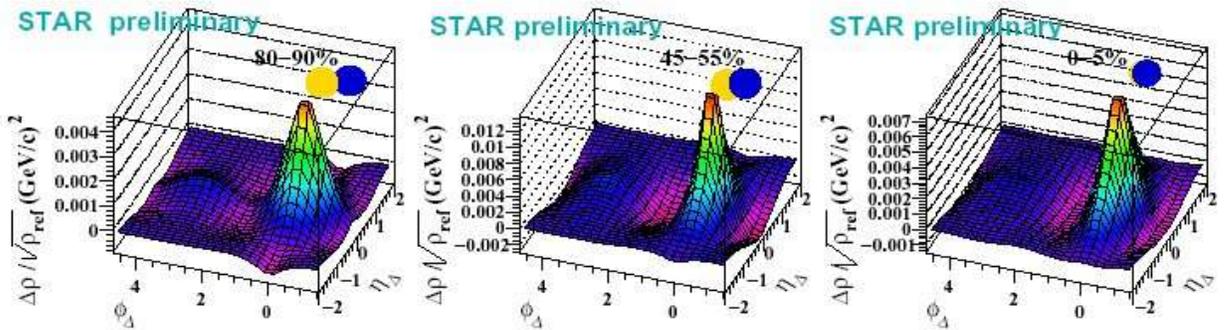

**Figure 5.3** Same data as in Fig.5.2 with subtracted azimuthal harmonics $v_1$ and $v_2$.

## Appendix A. Notation

In our notation $i$ is particle index, $n$ is number of particles within a kinematic cut

(bin), overline denotes an average over events, $\hat{p}_t$ is an inclusive mean $p_t$ per particle, $x_\Delta$ (variants: $\eta_\Delta$, $\phi_\Delta$) is difference variable $x_i - x_i'$, $\delta x$ is scale (range of local integration, see Fig.5.1), $\Delta x$ is the upper limit on $\delta x$. Covariance of $x$ and $y$ is denoted as $Cov[x,y]$, variance of $x$ -- as $Var[x]$.

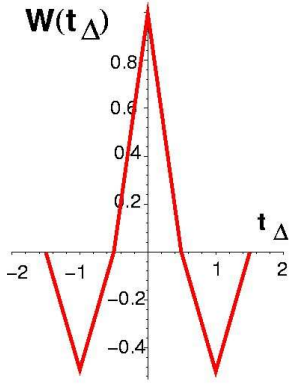

**Figure B.1** The integration kernel from Eq. B.6 for the case of a Haar wavelet, for $m=0$. Larger values of m will cause the function to shrink horizontally in factors of $2^m$, probing finer scales. Because of the specifics of its shape, an application of this kernel can be thought of as a comparison of an effective correlation content near 0 and away from 0, and thus as a differentiation or a shape-gradient measure. This is further illustrated by Fig.B.2.

## Appendix B. From "dynamic texture" to correlation

This connection is important to interpret the DWT-based analysis results (such as reported in [2]) in terms of correlation functions. Here I consider the case of a random function $X(t)$. The discrete-wavelet-based estimator of the power-spectrum on scale $m$ for the $X(t)$ is the normalized sum of the squared inner products between $X(t)$ and the discrete wavelet basis functions $f_{m,l}(t)$ ($l$ is location index):

$$P(m) = 2^{-m} \sum_l \overline{\left(\int X(t) f_{m,l}(t) dt\right)^2} \tag{B.1}$$

Equivalently,

$$P(m) = 2^{-m} \sum_l \overline{\int\int X(t_1) X(t_2) f_{m,l}(t_1) f_{m,l}(t_2) dt_1 dt_2} \tag{B.2}$$

We make a transition to the sum and difference variables $t_\Sigma$ and $t_\Delta$, preparing to consider the special case when the $t_\Sigma$ dependence can be neglected:

$$t_\Sigma = t_1 + t_2, \quad t_\Delta = t_1 - t_2 \tag{B.3}$$

$$t_1 = \frac{t_\Sigma + t_\Delta}{2}, \quad t_2 = \frac{t_\Sigma - t_\Delta}{2} \tag{B.4}$$

$$P(m) = 2^{-m-1} \sum_l \int\int \overline{X\left(\frac{t_\Sigma - t_\Delta}{2}\right) X\left(\frac{t_\Sigma + t_\Delta}{2}\right) f_{m,l}\left(\frac{t_\Sigma - t_\Delta}{2}\right) f_{m,l}\left(\frac{t_\Sigma + t_\Delta}{2}\right)} dt_\Sigma dt_\Delta \tag{B.5}$$

If the $t_\Sigma$ dependence of $X$ is indeed absent *on average*, in what follows we can set $t_\Sigma$ to any fixed value (such as 0) and factor $X$ out of the $t_\Sigma$ integral:

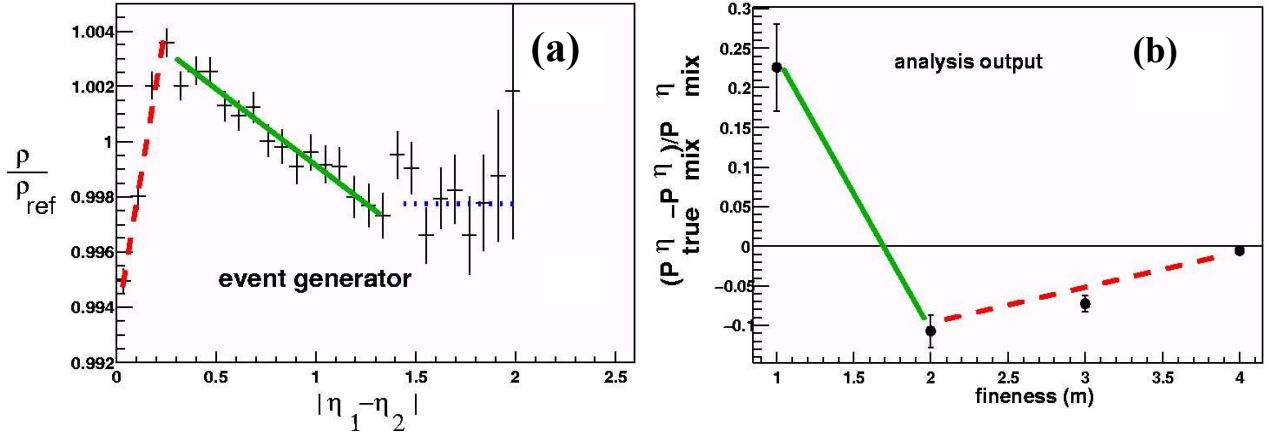

**Figure B.2. a)** A correlation function built into computationally synthesized events and reconstructed by direct analysis **b)** "dynamic texture" analysis output of the same event sample.

$$P(m) = 2^{-m-1} \overline{\int X\left(\frac{-t_\Delta}{2}\right) X\left(\frac{t_\Delta}{2}\right) \sum_l \left[\int f_{m,l}\left(\frac{t_\Sigma - t_\Delta}{2}\right) f_{m,l}\left(\frac{t_\Sigma + t_\Delta}{2}\right) dt_\Sigma\right] dt_\Delta} \qquad (B.6)$$

In general, $f_{m,l}(t) = 2^{m/2} F(2^m t - l)$ and for the Haar wavelet, $F(t) = G(2t) - G(2t-1)$, where $G(t) = 1$ when $0 \leq t \leq 1$ and 0 otherwise. The integral in the square brackets is a piece-wise-linear function $W(2^m t_\Delta)$, shown in Fig.B.1 for $m=0$. Equation B.6 furnishes the required connection between an *autocorrelation* (Equation C.1) and the power spectrum estimator $P(m)$, and thus between the "dynamic texture" $P_{dyn}(m) = P_{true}(m) - P_{mix}(m)$ and the reference-subtracted autocorrelation $\Delta\rho(t_\Delta)$. This method of characterizing the correlation shape is attractive because, first, it does not require one to sample particle pairs and is therefore fast. Second, the characterization of shapes is carried out using a set of quantities whose validity does not change from one application to another, thus facilitating the comparison of different physical systems – an alternative to the use of fitting functions whose parameters are specific to the particular model, often in need of justification.

**Appendix C. From variance to correlation**

The connection between variance and correlation function – the basis of analysis in Section 5 -- is discussed in detail using integral calculus in [5] and the algebra of discrete bins in [6]. For a random field $X(t)$, the *autocorrelation* function is defined [8] as

$$\rho(X,t_\Delta) \equiv \overline{X(t)X(t+t_\Delta)} \qquad (C.1)$$

The variance of $X$, integrated over bin of width $\delta\eta, \delta\phi$, centered at 0 (see Fig.5.1) is

$$Var[X;\delta\eta,\delta\phi] = \overline{(\int_{-\delta\eta/2}^{\delta\eta/2}\int_{-\delta\phi/2}^{\delta\phi/2} X\, d\eta\, d\phi)^2} - (\int_{-\delta\eta/2}^{\delta\eta/2}\overline{X}\, d\eta\, d\phi)^2 = \qquad (C.2)$$

$$\int_{-\delta\eta/2}^{\delta\eta/2} d\eta_1 \int_{-\delta\phi/2}^{\delta\phi/2} d\phi_1 \int_{-\delta\eta/2}^{\delta\eta/2} d\eta_2 \int_{-\delta\phi/2}^{\delta\phi/2} d\phi_2 [\overline{X(\eta_1,\phi_1)X(\eta_2,\phi_2)} - \overline{X(\eta_1,\phi_1)}\,\overline{X(\eta_2,\phi_2)}] \qquad (C.3)$$

As we form the reference-subtracted variance

$$\Delta\sigma_X^2 \equiv Var[X;\delta\eta,\delta\phi] - Var[X_{ref};\delta\eta,\delta\phi], \qquad (C.4)$$

the first term in the brackets above becomes the reference-subtracted correlation $\Delta\rho(X,\eta_\Delta,\phi_\Delta)$ and the second term cancels with the reference since $\overline{X} = \overline{X_{ref}}$.

$$\Delta\sigma_X^2 = \int_{-\delta\eta/2}^{\delta\eta/2} d\eta_1 \int_{-\delta\phi/2}^{\delta\phi/2} d\phi_1 \int_{-\delta\eta/2}^{\delta\eta/2} d\eta_2 \int_{-\delta\phi/2}^{\delta\phi/2} d\phi_2 \Delta\rho(X,\eta_1-\eta_2,\phi_1-\phi_2) = \qquad (C.5)$$

$$2\int_0^{\delta\eta} d\eta_\Delta 2\int_0^{\delta\phi} d\phi_\Delta (\delta\eta-\eta_\Delta)(\delta\phi-\phi_\Delta)\Delta\rho(X,\eta_\Delta,\phi_\Delta) \qquad (C.6)$$

Equation C.6 is obtained by making a change of the integration variables from $\eta_1$, $\eta_2$ to $\eta_2$ and $\eta_\Delta = \eta_1 - \eta_2$, integrating over $\eta_1$, and repeating the same for $\phi$ (see [2] for details). In the actual analysis, this integral equation is discretized by partitioning the acceptance into $m_\delta \times n_\delta$ microbins of size $\epsilon_\eta \times \epsilon_\phi$. $X$ is replaced by $(p_t - n\hat{p}_t)/\sqrt{n}$. This denominator creates the reference number correlation $\rho_{ref}(n)$ in the final expression C.7 which makes it a per-particle measure:

$$\Delta\sigma_{p_t:n}^2(m_\delta\epsilon_\eta, n_\delta\epsilon_\phi) = 4\sum_{k,l=1}^{m_\delta,n_\delta} \epsilon_\eta\epsilon_\phi K_{m_\delta n_\delta : kl} \frac{\Delta\rho(p_t:n;k\epsilon_\eta,l\epsilon_\phi)}{\sqrt{\rho_{ref}(n;k\epsilon_\eta,l\epsilon_\phi)}} \qquad (C.7)$$

Here the kernel $K$ is a discrete representation of the continuous case:

$$(\delta\eta-\eta_\Delta)(\delta\phi-\phi_\Delta) \to \epsilon_\eta\epsilon_\phi K_{m_\delta n_\delta : kl} \equiv \epsilon_\eta\epsilon_\phi (m_\delta - k + \frac{1}{2})(n_\delta - l + \frac{1}{2}) \qquad (C.8)$$

Knowing the computationally cheaper quantity $\Delta\sigma_{p_t:n}^2$, we solve the integral equation B.7 for $\Delta\rho/\sqrt{\rho_{ref}}$ using standard numerical techniques[6].